\documentstyle[prl,preprint,amssymb,aps,]{revtex}
\newcommand{\xminus}[1]{\mbox{$X^-_{#1}$}}
\newcommand{\Alx}[2]{\mbox{Al$_{#1}$Ga$_{#2}$As}}
\draft
\begin{document}
%\wideabs{
\title{Internal transitions of negatively charged magneto-excitons
and many body effects in a two-dimensional electron gas}
\author{H.A.~Nickel, T.M.~Yeo, A.B.~Dzyubenko,\cite{ABD}
         B.D.~McCombe, and A.~Petrou}
\address{Department of Physics and Center for Advanced Photonic and
Electronic Materials, SUNY Buffalo, Buffalo, NY 14260, USA}
\author{A.Yu.~Sivachenko}
\address{Department of Physics, University of Utah, Salt Lake City,
UT 84112, USA}
\author{W.~Schaff}
\address{School of Electrical Engineering, Phillips Hall, Cornell University,
Ithaca, NY 14853, USA}
\author{V.~Umansky}
\address{Department of Condensed Matter Physics, The Weizmann Institute of Science,
76100 Rehovot, Israel}
\date{\today}
\maketitle
\begin{abstract}
Spin-singlet and spin-triplet internal transitions of
quasi-two-di\-men\-sio\-nal, ne\-ga\-tively charged ma\-gne\-to-exci\-tons
(\xminus{})
and their evolution with excess electron density
have been studied in GaAs/AlGaAs quantum-wells
by optically detected resonance (ODR) spectroscopy.
In the dilute electron limit, due to  magnetic translational invariance,
the ODR spectra are
dominated by bound-to-continuum bands in contrast to the superficially
similar negatively-charged-donor system $D^-$, which exhibits strictly
bound-to-bound transitions.
 With increasing excess electron density in the wells in
the magnetic field region corresponding to Landau level filling factors
$\nu < 2$ the \xminus{}-like transitions are {\em blue-shifted\/};
they are absent for $\nu > 2$.
The blue-shifted transitions are explained in terms of
a new type of collective excitation --- magnetoplasmons
bound to a mobile valence band hole, which demonstrates the
many-body nature of ``exciton-like'' magnetoluminescence
for $\nu < 2$.
\end{abstract}
%%%%%%%%%%%%%%%%%%%%%%%%%%%%%%%%%%%%%%%%%%%%%%%%%%%%%%%%%%%%%%%%%
%} % end of wideabs
%\begin{multicols}{2}

Since the initial observation of negatively charged excitons \xminus{}
in quantum wells (QWs) \cite{kheng93a}, there has been considerable
interest in {\em charged electron-hole complexes\/}
in quasi-two-dimensional (quasi-2D) semiconductor systems.
The negatively charged exciton, like its atomic physics analog, 
the negatively charged hydrogen ion $H^-$,
provides a central example of the role of 
{\em electron-electron correlations\/} 
in a Coulomb potential.  However, \xminus{} in QW structures
is different in several important respects: (1) the positively charged
holes have a mass comparable to the electron mass; (2) the magnetic field
energy can dominate at laboratory magnetic
fields; (3) the system is quasi-2D; and (4) the density
of excess electrons $N_e$ in the wells can be controlled (at low $N_e$,
\xminus{} is the {\em stable ground state} of the photoexcited carrier
system).
Additionally, in contrast to one-component electron systems with
free carriers or carriers in a parabolic confining potential
\cite{kohn61a,dempsey92a}, the corrections to the single-particle
states are directly measurable by {\em intraband\/} optical experiments.

The evolution with $N_e$ of the electron-hole ($e$--$h$) system in
quasi-2D structures from isolated neutral excitons via \xminus{}
to a few-hole/many-electron plasma, and the effects of a magnetic
field have been studied by interband methods
\cite{finkelstein95a,shields95a,gekhtman96a,hawrylak97,cox98a,rashba00}.
At zero field the \xminus{}-feature  evolves with increasing $N_e$
into the Fermi-edge of the $e$--$h$ plasma \cite{huard00}; the density
at which the crossover takes place depends on the inherent disorder
in the sample \cite{finkelstein95a}.  When the electrons and holes
are confined in the same spatial region along the growth direction,
the magneto-photoluminescence (PL) or magneto-absorption 
changes abruptly with increasing
magnetic field at Landau-level (LL) filling factor $\nu = 2$ from bandgap
renormalized LL-to-LL-like transitions ($\nu > 2$) to ``exciton-like''
behavior ($\nu < 2$) irrespective of $N_e$.  
The nature of the ``exciton-like'' states for $\nu <2$ remains not
well understood \cite{gekhtman96a}, and the {\em interband\/} measurements
alone can not distinguish between a collective, many-body state and
a dilute $X^-$ system (see \cite{hawrylak97,rashba00} and references therein).
Internal excitonic transitions (IETs), which are now understood for neutral
\cite{groeneveld94a,odrpapers1} and charged \cite{dzyubenko00a} excitons,
probe directly the ground and excited state properties,
and thus can be used for investigating the excitonic  
state in the dilute situation and its
evolution with $N_e$ and magnetic field.

The \xminus{}-complex, although superficially similar to the negatively
charged donor $D^-$ \cite{dzyubenko93a,jiang97a}, differs in one
very important respect: the positive charge in \xminus{} is mobile.
A symmetry associated with the resulting center-of-mass (CM) motion
(magnetic translational invariance) leads to a new, exact electric-dipole
selection rule that {\em prohibits} internal transitions between certain
families of \xminus{} states \cite{dzyubenko00c}.  In particular, the
spin-singlet and -triplet bound-to-bound transitions that dominate the
spectra of $D^-$ \cite{dzyubenko93a,jiang97a} are {\em strictly
forbidden} for \xminus{}.
The present experimental observation and studies of internal
transitions of \xminus{} in GaAs QWs as functions of $N_e$
and magnetic field provide new insight into this complex system.
At low $N_e$ bound-to-bound transitions are
absent, and both singlet and triplet features appear as continuous bands,
with positions in quantitative agreement with numerical calculations.
At high $N_e$ and low fields ($\nu > 2$) {\em internal transitions are not
observed}; in contrast, for $\nu < 2$ they appear but are {\em blue-shifted\/}
in energy from their low-$N_e$ counterparts.  These experiments
(1) verify the predicted consequences of the magnetic translational symmetry
\cite{dzyubenko00c}, and (2) more importantly, clearly show
that the feature identified as \xminus{} for $\nu <2$ 
represents the {\em collective response}
of a few-hole/many-electron system.

The exact selection rules for IETs of isolated charged
excitons \cite{dzyubenko00c} are shown in Fig.~1a.
In the strictly-2D high-field limit the only family
(of macroscopically degenerate) bound states
associated with the lowest $n_e=0$, $n_h=0$ LL
is the triplet \xminus{t00} \cite{wojs95,palacios96a,whittaker97a}.
The \xminus{} states are properly
specified in magnetic fields \cite{dzyubenko00c}
by the total electron spin $S_e$ ($S_e=0$
for singlet, $s$, and $S_e=1$ for triplet, $t$, states), by the hole spin
$S_h$ ($S_{\rm hz} = \pm 3/2$ for heavy-holes), by the total angular
momentum projection $m_z$, and by the discrete oscillator quantum
number $k=0, 1, \ldots$ (related to the center of
rotation of the complex). Due to macroscopic LL degeneracy in $k$,
\xminus{} states form degenerate families.  Each family starts with
its $k=0$ parent state (solid dots in Fig.~1a),
which has a specific value of
angular momentum projection, $m_z^{k=0}$,
determined by particulars of the interactions.
The degenerate daughter states
$k=1,2, \ldots$ have angular momentum projections $m_z=m_z^{k=0}-k$
(open dots in Fig.~1a).
In the next
electron LL ($n_e=1$, $n_h=0$), there is also only one (macroscopically
degenerate) bound triplet state, \xminus{t10} \cite{dzyubenko00c}.
In contrast to $D^-$ \cite{dzyubenko93a},
the \xminus{} eigenstates consist of discrete bound states {\em and} continua
(due to the extended CM motion of the neutral magneto-exciton) and, in general,
the FIR spectra include both bound-to-bound and bound-to-continuum
transitions.  The bound-to-bound transition
$\xminus{t00} \rightarrow \xminus{t10}$
(dotted arrow in Fig.~1a)
is allowed by the usual selection rules
($m_z \rightarrow m_z + 1$ for $\sigma^+$ polarization, spin conserved).
This is the analog of the strong $T^-$-transition for the $D^-$
center \cite{dzyubenko93a,jiang97a}; it lies {\em below}
the electron cyclotron resonance (e-CR) in energy.
For \xminus{}, however, the {\em additional exact selection rule\/}
$\Delta k = 0$ must also be satisfied, reflecting
the fact that the center of rotation of the complex in ${\bf B}$
cannot be displaced by the absorption of a photon propagating
along the field direction.
Because the CM and internal motions are coupled in ${\bf B}$, 
$k$ and $m_z$ are not independent,
and {\em both\/} selection rules cannot be satisfied simultaneously for
the bound-to-bound transition, $\xminus{t00} \rightarrow \xminus{t10}$.
The oscillator strength is transferred to bound-to-continuum transitions
($T_1$ and $T_2$ in Fig.~1a). In addition to a magnetoexciton band of width
$E_0 = \sqrt{\pi/2} \, e^2/\epsilon l_B$,  $l_B = (\hbar c /eB)^{1/2}$,
extending below each LL and describing a $1s$ exciton plus a
scattered electron in the corresponding LL \
($X_{00} + e_0$ and $X_{00} + e_{1}$),
the continuum of three particle states (hatched regions) also includes
a band of width $0.57 E_0$ below the first excited electron LL. The
latter corresponds to an excited ($2p^+$) exciton plus a scattered
electron in the zero LL ($X_{10} + e_0$).
Transitions to the $X_{00} + e_1$ continuum are dominated by a sharp onset
at the edge (energy of e-CR plus the \xminus{} binding energy:
solid arrow $T_1$ in Fig.~1a (see also Fig.~1b).
In addition, there is a broader,
weaker peak associated with transitions to the lower edge of the
$X_{10} + e_0$ magnetoexciton band (dashed arrow $T_2$).
Detailed calculations show that these qualitative triplet features
remain at finite fields and confinement,
{\em and also hold for the singlet \xminus{} states\/}, which form
the ground state \cite{whittaker97a}
at low and intermediate fields.
The singlet counterparts of the triplet transitions
$T_1$ and $T_2$ (see Fig.~1a) are denoted below as $S_1$ and $S_2$.

Four GaAs/\Alx{0.3}{0.7} multiple~quantum~well
(MQW) samples grown by molecular beam epitaxy
were investigated.  Samples 1 and 2 are nominally undoped with
(20\,nm) GaAs wells and \Alx{0.3}{0.7} barriers:
Sample 1 -- 20 wells with 60\,nm barriers; Sample 2 --
10 wells with 20\,nm barriers. Samples 3 and 4 have 24\,nm wells and are
Si-doped in the central third
of the \Alx{0.3}{0.7} barriers at sheet densities of $8 \times 10^{10}$
and $2.8 \times 10^{11}$ cm$^{-2}$, respectively: Sample 3 (4) -- 20 (10)
wells with 48\,nm (24\,nm) barriers.
The FIR absorption resonances in a magnetic field were studied by
ODR spectroscopy \cite{odrpapers1}.
Electrons, holes, and excitons were continuously created by optical
excitation with the 632.8\,nm line of a He-Ne laser coupled to the sample
(at low temperature in the Faraday geometry in the center of a 15/17 T
superconducting magnet) through an optical fiber; PL was collected
via a second fiber \cite{kono95a}.
Photo-excited exciton densities are estimated to be approximately
$10^7\,$cm$^{-2}$ at an excitation laser intensity of 100\,mW/cm$^2$.

Raw ODR data for sample~1 are shown in Fig.~2a;
Fig.~3a shows a summary of the triplet and singlet
ODR data from the two undoped samples 1 and 2.
The ODR signal obtained
by tracking the \xminus{}-PL feature (Fig.~3b) is plotted vs magnetic field
at several FIR laser photon energies. Very similar, but
positive-going, ODR spectra are obtained by tracking the $X$-PL line.
The sharp feature at 6.3\,T at 10.4 meV is e-CR.
Three features are
attributed to \xminus{} internal transitions ($S_1$, $S_2$, and $T_1$);
similar features are seen in sample~2.
The weak shoulder at fields just
below e-CR is ascribed to the dominant triplet ionizing transition $T_1$.
The peak of this band corresponds to the edge of the continuum,
so the observed peak is shifted to higher energies from
e-CR by the (small) triplet binding energy, with a tail at
higher energies (lower fields). 
Quantitative numerical calculations
for a 20\,nm GaAs QW performed
using an expansion in LL's incorporating size quantization \cite{dzyubenko93a},
are shown in Fig.~3a at 6 and 9\,T by the circles with crosses.
We estimate the absolute accuracy achieved
in determining the \xminus{} binding energies to be $0.1 - 0.2$\,meV
with no adjustable parameters\cite{numcalcnote}.  The agreement
with experiment is excellent and provides quantitative support for the
assignment of the observed features \cite{dzyubenko00a}.

The clearest signature of the new physics associated with the
magnetic translational invariance is the $X^-$-triplet transition.
The bound-to-bound transition (dotted arrow in Fig.~1a)
should occur on the {\em high-field\/} (low energy) side of the e-CR.
There is {\em no indication\/}
of a line at this position in the ODR data of Fig.~2.
The strongest feature seen near e-CR is a broad band on the
{\em low-field\/} side, precisely as predicted for the allowed
and strong \cite{dzyubenko00c}
bound-to-continuum triplet transition of $X^-$, $T_1$.

To explore the effects of large densities of excess electrons
and the interesting behavior of the magneto-PL above and below $\nu=2$
we have also investigated modulation-doped samples under very 
similar conditions. ODR results for sample~3
($N_e=8 \times 10^{10}$ cm$^{-2}$) are shown in Fig.~2b; 
in this case the only PL feature seen at low temperature and 
low fields evolves into an \xminus{}-like line for $\nu < 2$ (Fig.~3b).
It is clear from Fig.~2b that for laser photon energies
of 6.73\,meV and above,  the ODR band just below the sharp e-CR 
(the triplet band $T_1$\/) is now much stronger than it is for 
low $N_e$ and is also shifted to lower magnetic fields
(a blue shift in energy).
The large width of this band is also apparent.
At this density, filling factor $\nu = 2$ occurs at
$B \simeq 1.6$\,T. At smaller photon energies,
for which the resonant field would lie below that of $\nu=2$,
there is no evidence of this triplet band, and the e-CR
line is broader and symmetric.
Note also that the {\em dominant\/}  singlet-like line $S_2$,
which at low $N_e$ corresponds
to the $X^-_s \rightarrow X_{10} + e_0$ transition,
is clearly seen at the two highest laser photon energies, for
which the resonance occurs at fields corresponding to $\nu < 2$.
This feature loses strength at a photon energy of 10.4 meV,
and is not observable at 6.73 meV
and below (resonant fields corresponding to $\nu > 2$).

The data for sample~4 are similar, except that due to the higher density,
the singlet-like $S_2$ and
triplet-like $T_1$ bands are observable only at fields above 5\,T
($\nu$ = 2 at 5.8\,T).
Features observed in samples~3 and 4 occur at lower fields than the
corresponding features in the nominally undoped samples --- they are
{\em blue-shifted} in energy.  This measured blue shift of the
dominant \xminus{} feature $S_2$
is shown in Fig.~3d  along with data \cite{tischler} for the
$D^-$ singlet blue shift \cite{boundMPL}
for a well-- and barrier--doped sample
of approximately equal dimensions and $N_e$.

The blue shift is the signature of a many-electron phenomenon; the relevant
collective excitations are magnetoplasmons.
In the absence of the valence band hole, they are characterized
by the conserved center-of-mass momentum ${\bf K}$ and are
neutral bound $e$--$h$ pairs \cite{bychkov81}.
Due to Kohn's theorem \cite{kohn61a}, only ${\bf K}=0$
magnetoplasmons with the bare e-CR energy
$\hbar\omega_{\rm ce}$ are FIR-active.
In the presence of the valence band hole,
the final states are {\em charged mobile\/} $e$--$h$ complexes
in a magnetic field \cite{hawrylak97,dzyubenko00c} (see Fig.~1c).
The interaction
with the valence band hole mixes various ${\bf K}$-states, obviating
Kohn's theorem; many states acquire oscillator
strength. Our calculations for 2D systems with integer filling factors
$\nu =1,2$ show that there is one prominent absorption peak
in the region of energies larger than $\hbar\omega_{\rm ce}$; it corresponds
to a {\em magnetoplasmon bound to the mobile hole\/} $h^+$.
These excitations resemble the magnetoplasma modes bound
to the {\em fixed\/} donor ion $D^+$ in the presence of many
electrons \cite{boundMPL}; but
the magnetic translational invariance is broken for the latter.
Energies of bound collective modes
experience discontinuities at integer $\nu$
(cf.\ \cite{boundMPL,hawrylak97})
and increase with increasing $\nu$, reflecting the enhanced
contribution of the exchange-correlation effects and explaining
the blue shift.
As seen in Fig.~3d the measured blue shifts
for the singlet \xminus{} with the {\em mobile} hole are, somewhat surprisingly,
{\em larger\/} than those for the $D^-$ for fields above 5\,T.
This is in qualitative agreement with theory for the
blue shift of the $X^-$ triplet internal transition
(see Fig.~3c). The larger blue shift for the $X^-$ results
from the diminished negative contribution of the Coulomb $e$--$h$ interaction
to the final state energy for the mobile hole.
We conclude that ``$X^-$-like'' PL features
seen in QWs with excess electrons at $\nu < 2$ represent a
collective response of the electron-hole system,
which gradually approaches the isolated $X^-$ for $\nu \ll 1$.

Work supported in part by NSF DMR 9722625 and by a COBASE grant.
We are grateful to the CCR at UB for providing the supercomputer facilities.

%%%%%%%%%%%%%%%%%%%%%%%%%%%%%%%%%%%%%%%%%

%%%%%%%%%%%%%  FIGURES  %%%%%%%%%%%%%%%%%%%%%%%%%%%%%

\begin{figure}
\caption{Schematic diagram of
(a) the $X^-$ triplet energy states associated
with the two lowest electron Landau
levels for the strictly 2D, high magnetic field limit.
Initial and final states in the FIR-transition
(b) for the isolated $X^-$ and
(c)  for electron filling factor $\nu = 1$.}
		\label{f.summaryplot}
\end{figure}

\begin{figure}
\caption{ODR spectra for (a) sample~1 and (b) sample~3.
The features labeled $S_1$, $S_2$, and $T_1$
refer to the internal singlet and triplet transitions of \xminus{} discussed
in the text.}
		\label{f.rawdata}
\end{figure}

\begin{figure}
\caption{(a): Summary plot for samples~1 (closed symbols) and 2 
(open symbols): internal singlet $S_1$ ($\blacktriangle, \vartriangle$),
$S_2$ ($\blacksquare, \square$),
and triplet $T_1$ ($\blacktriangledown, \triangledown$)
transitions of $X^-$. Open circles with crosses:
numerical calculations. Dashed lines are guides to the eye.
(b) Magneto-PL for sample 3 from the
ground and excited $X^-$ states ($\small{ \blacksquare}$):
At $B \sim 7$~T the two different hole spin states are resolved.
The highest energy line ($\bullet$) above 6~T is the neutral exciton, $X$.
Bandgap renormalized LL-to-LL transitions (linear in $B$)
are seen below 1.8~T.  Up arrows denote the expected field positions
for the $S_2$ transition at the indicated photon energies (Fig.~2b).
(c) Calculated energies
(in units of $E_0 = \protect\sqrt{\pi/2} \, e^2/\epsilon l_B$)
of magnetoplasmons bound to
$D^+$ and $h^+$. The heavy solid and dotted lines indicate the
transition energies for the
isolated, strictly-2D $D^-_{t,s}$ and $X^-_t$ states, respectively.
Blue-shifts are indicated by double-headed arrows.
(d) Experimental blue-shift in \xminus{} singlet-like ($\blacktriangle$)
and $D^-$ singlet-like ($\blacktriangledown$) \protect\cite{tischler}
transitions.}
		\label{f.blueshift}
\end{figure}


\begin{references}
\bibitem{numcalcnote}
An iteration procedure similar to that of \cite{whittaker97a} was used for
picking out a relatively large but limited subspace (several thousand basis
elements) of the total Hilbert space, so as to maximize the projection
of the true eigenstate.
The following parameters for the 20~nm GaAs/\Alx{0.7}{0.3} QW were used:
isotropic effective electron mass $m_e=0.067$, in-plane
(perpendicular) hole mass $m_{h||}=0.24$ ($m_{h\bot}=0.34$), dielectric
constant $\epsilon=12.5$, total band-edge discontinuity $\Delta U =
\Delta U_e + \Delta U_h = 352$ meV with a ratio $\Delta U_e/\Delta
U_h=70:30$.
\end{references}
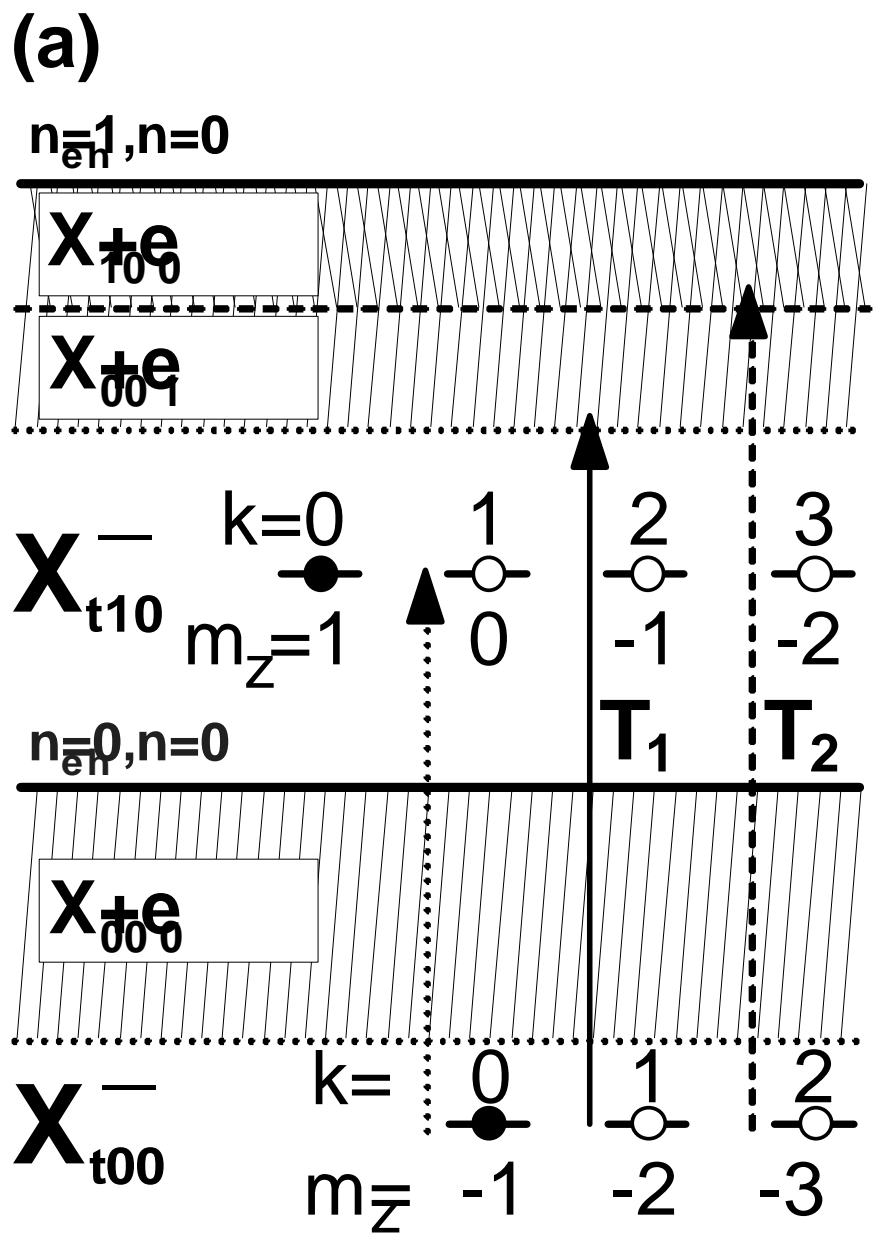
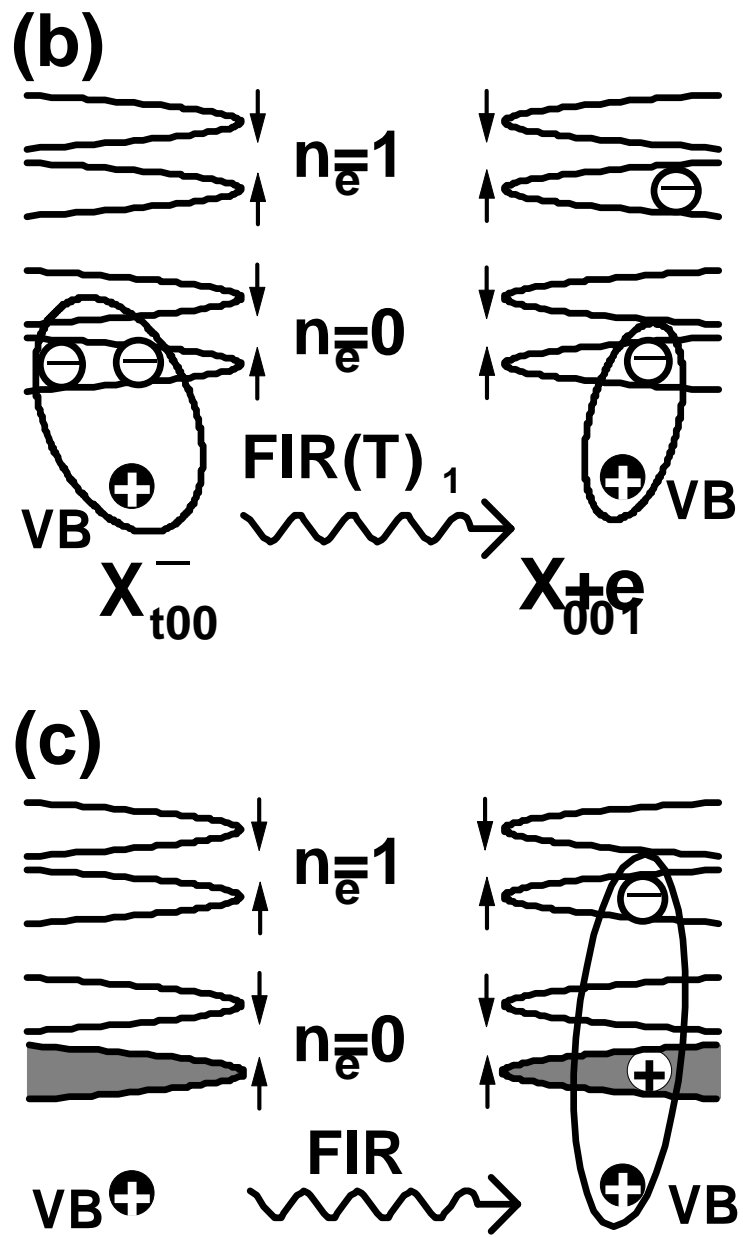
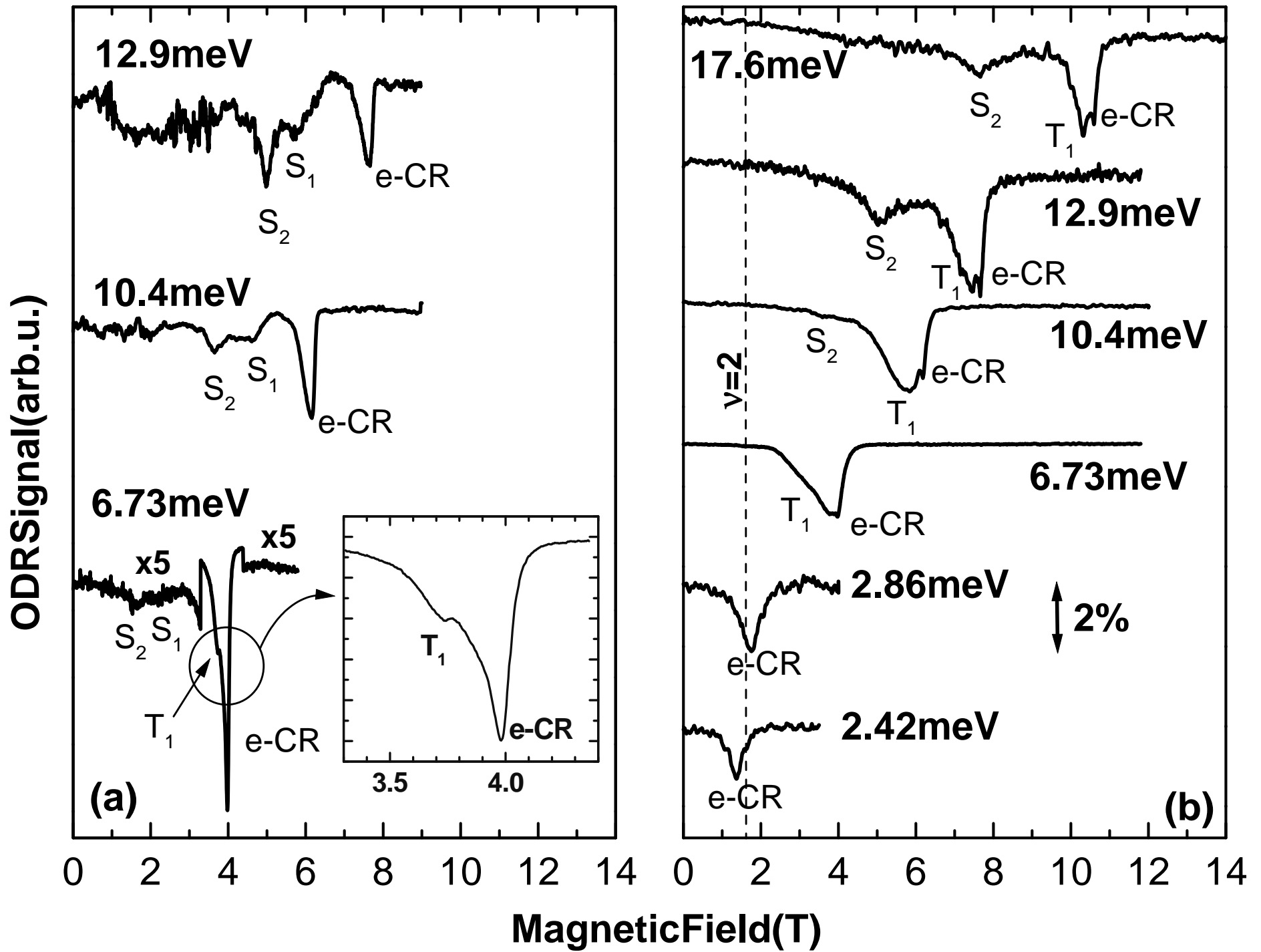
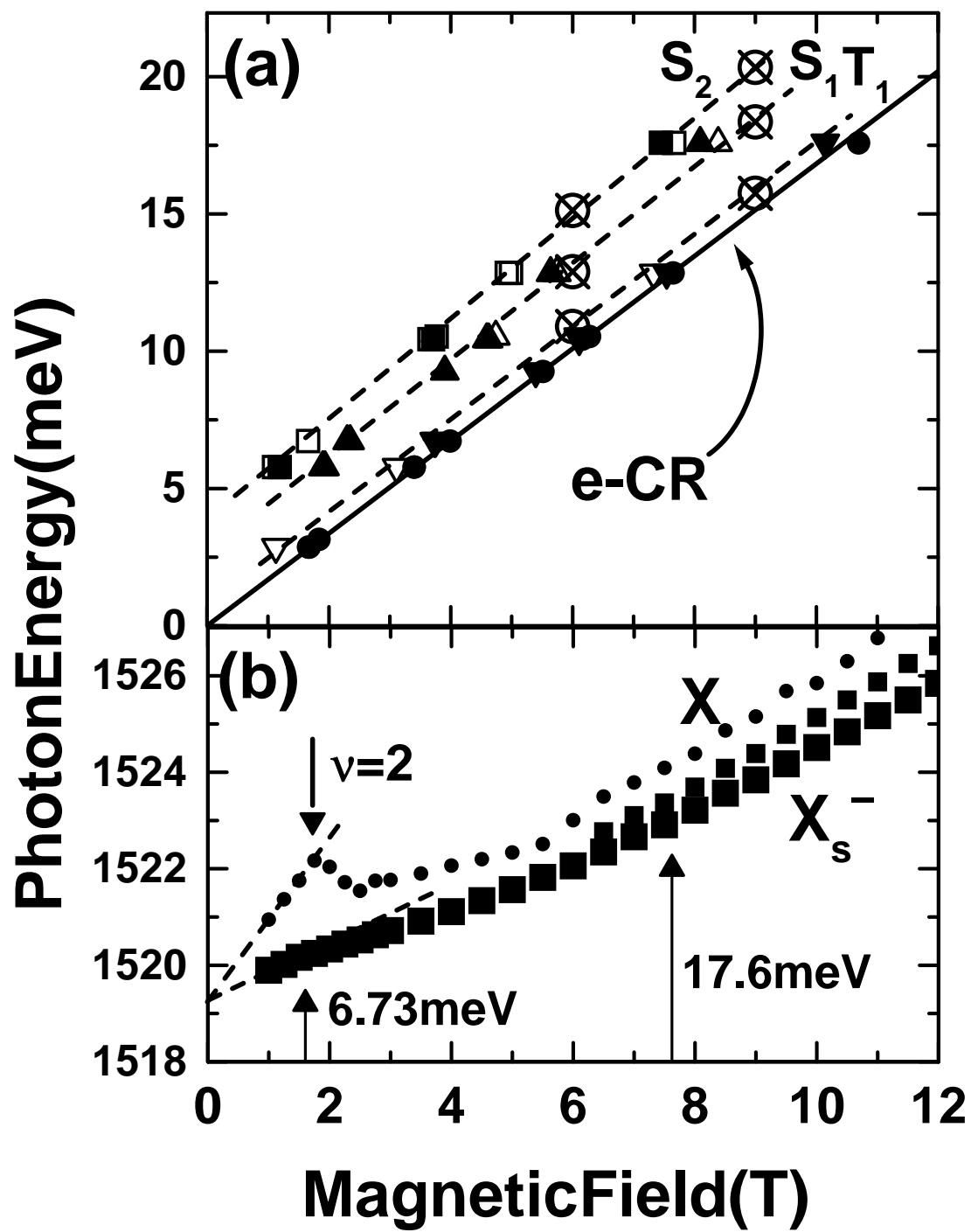
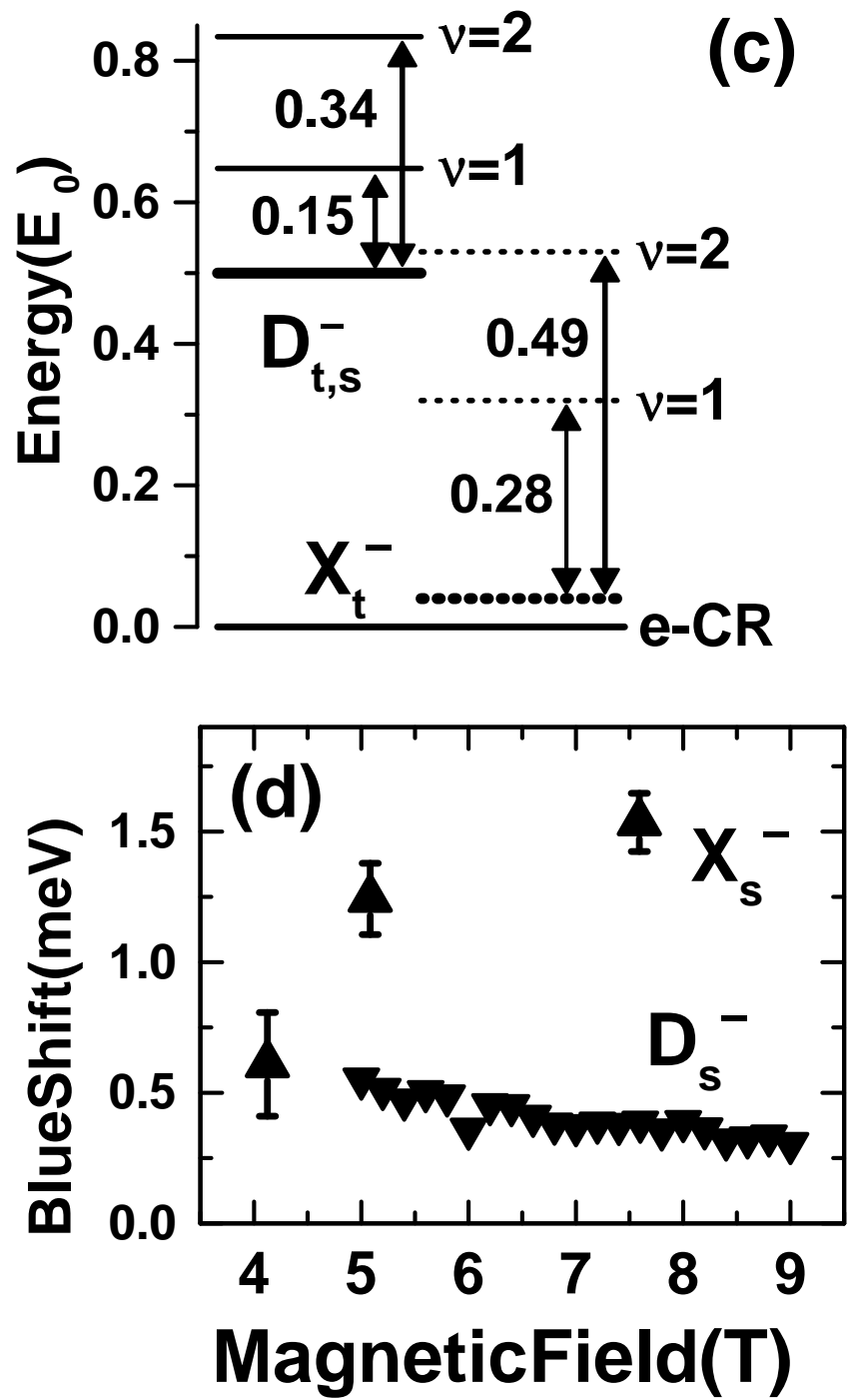
\end{document}